\documentstyle[twocolumn,prl,aps,epsfig]{revtex}
\newcommand{\beq}{\begin{eqnarray}}
\newcommand{\eeq}{\end{eqnarray}}
\renewcommand{\vec}[1]{{\mathbf{#1}}}
\begin{document}
\draft

\title
{Magneto-transport Properties near the Superconductor-Insulator
Transition in 2D}
\author{Denis Dalidovich and Philip Phillips}\vspace{.05in}

%
\address
{Loomis Laboratory of Physics\\
University of Illinois at Urbana-Champaign\\
1100 W.Green St., Urbana, IL, 61801-3080}

%

\address{\mbox{ }}
\address{\parbox{14.5cm}{\rm \mbox{ }\mbox{ }
We analyze here the behavior near the 2D insulator-superconductor
quantum critical point in the presence of a perpendicular
magnetic field. We show that with increasing field $H$, the quantum
disordered and quantum critical regimes, in which vortex degrees of
freedom are suppressed, crossover to a new
magnetically activated (MA) regime, where the correlation
length $\xi \sim 1/\sqrt{H}$. In this regime, we show that the conductivity
decreases monotonically as opposed to the anticipated saturation predicted
from hyperuniversality arguments.  This discrepancy arises from the lack of
commutativity of the frequency and temperature tending to zero limits of the
conductivity. In the low-field regime
such that $\sqrt{H}\ll\Delta$, and in the absence of Ohmic dissipation,
where $\Delta$ is a measure of the distance from the quantum critical
point, the resistivity saturates to the Bose metal value found previously
for Cooper pairs lacking phase coherence.
}}
\address{\mbox{ }}
\address{\mbox{ }}

\maketitle

From the earliest experiments\cite{hp,yk,gm} on superconducting
thin metal alloy films, a simple picture emerged for the magnetic-field
tuned transition to the insulating state. Namely, the transition
is continuous with a well-defined critical field, $H_c$, below which
superconductivity occurs and above which an insulator obtains.
At the critical field, the resistivity is independent of
temperature.  Within the dirty boson model\cite{mf} in an
applied magnetic field, these results follow naturally as a direct
consequence of the inherent duality between Cooper pairs and
vortices:  above $H_c$, vortices condense while Cooper pairs remain
localized but below $H_c$ it is the Cooper pair condensation and
vortex localization that affords the zero-resistance state.  At criticality,
Cooper pairs on the brink of forming a phase coherent
state diffuse with the quantum of conductance $\sigma_Q=4e^2/h$ for
charge $2e$ bosons. However, more recent experiments have shown that
this simple picture breaks down dramatically at and below $H_c$.
In particular, a metallic phase\cite{mk,ephron} emerges below $H_c$ with the
zero-resistance state occurring only at a much
smaller field $H_M\ll H_c$\cite{mk2}. On phenomenological
grounds, it has been suggested that the metallic phase arises
from the quantum tunneling of field-induced vortices in
the presence of dissipation\cite{shimshoni}.

As the metallic phase has been observed at zero-field\cite{jaeger} in
the disorder-tuned insulator-superconductor transition as well,
we reinvestigated the conductivity of the standard quantum-disordered
regime in zero field. We found that quite generally
Cooper pairs lacking phase coherence form a Bose metal phase\cite{denis2}.
This result arises from the fact that in the quantum-disordered regime the
elementary excitations obey Boltzmann statistics. Hence, their 
concentration $n$ is exponentially suppressed. As a consequence, 
so is the inverse scattering 
time $1/\tau$ that is determined by the probability that
two quasiparticles collide. But because the conductivity is a product of the 
density and the scattering time, the exponentials cancel one another, thereby
leading to a finite conductivity as $T\rightarrow 0$.
In the current study,
we focus on the fate of the Bose metal phase in a
magnetic field.   Near the quantum critical point where vortex dynamics
are subdominant, the relevant
physics can be accessed by the standard $O(M)$ Ginzburg-Landau (GL)
theory\cite{cha}. In this work, we utilize this approach for
a system near the IST in the presence of a perpendicular magnetic field.
We establish that at low temperature $T$ and close to the quantum
critical point, $\Delta\ll 1$, the application of a field $H$ drives the system
to the high-field {\it magnetically activated} (MA) regime
once $\sqrt{H} \gg \Delta, T$, thereby compounding the analysis of the
transport properties in the quantum disordered regime.
In this regime, the energy levels of the bosonic quasiparticle excitations
are quantized and have a gap $m \sim \sqrt{H}$ that is inversely
proportional to the correlation length $\xi$.  Working within the
large $M$ approximation, we discuss the crossovers and
limiting forms of the scaling functions for the correlation length
$\xi=\xi(H,T,\Delta)$ and the static conductivity $\sigma$
in the vicinity of the critical point. We present the qualitative form for
the conductivity in the presence of dissipation based on the standard Kubo
formula neglecting the effects of frustration.
The dc conductivity is regularized by inclusion of the weak coupling to
an Ohmic heat bath $\eta$ as well by quasiparticle scattering $1/\tau$ due to
the quartic non-linearity in the GL action. 
Collectively, internal and external dissipation can be treated
jointly with an effective coupling constant ${\tilde \eta}=\eta+1/\tau$.
In the absence of the bath and at a
small constant magnetic field such that $\sqrt{H}\ll\Delta$, the magnetic
energy scale is irrelevant.  Consequently, the Bose metal\cite{denis2}
obtained previously
in the quantum disordered regime is recovered.
We show that in the MA regime, the conductivity $\sigma$ is directly,
rather than inversely, proportional to the total dissipation
${\tilde \eta}$. Further, in contrast to
the low-field limit, the conductivity in the MA regime
is found to be a decreasing function of
temperature and inversely related to the field.

Our starting point
is the $M$-component Ginzburg-Landau free energy\cite{ds,denis},
\beq\label{action}
F[\psi]&=&\int d^2r\int d\tau\left\{
\left[\left(\nabla+\frac{ie^*}{\hbar}\vec A_0(\vec r)\right)
\psi_a^*(\vec r,\tau)\right]\right.\nonumber\\
&&\left.\cdot
\left[\left(\nabla-\frac{ie^*}{\hbar}\vec
A_0(\vec r)\right)\psi_a(\vec r, \tau) \right]
+ \left|\partial_\tau\psi_a(\vec r,\tau)\right|^2
\right.\nonumber\\
&&\left.
+\delta \left|\psi_a(\vec r,\tau)\right|^2 +
\frac{U}{2M}\left|\psi_a (\vec r,\tau)\right|^4 \right\},
\eeq
in the presence of a magnetic field,
where $\vec A_0(\vec r)=\{ 0, Hx, 0 \}$ is the vector potential due to the
applied field, $e^*=2e$, and $\delta$ is the $T=0$ unrenormalized distance to
the quantum critical point. For a time-independent magnetic field, it is
convenient to expand the order parameter $\psi_a(\vec r,\tau)$ in the
corresponding eigenfunctions
that diagonalize the Gaussian part of the free-energy density
\beq\label{psi}
\psi_a(\vec r,\tau)=\sum_{\omega_m , p_y, n}
e^{-ip_y y-i\omega_m \tau} \Psi_n (x-\frac{\hbar p_y}{e^* H})
c_a(n, \omega_m, p_y)
\eeq
where $\Psi_n (x-\hbar p_y/e^* H)$ is the solution of the corresponding
eigenequation, expressible in Hermite polynomials,
and $c_a(n, \omega_m, p_y)$ are the expansion coefficients. The related
eigenvalues are given by $\omega_{H}(n+1/2)$ with
$\omega_{H}=2e^* Ha/\hbar =4\pi f/a$, where $a$ is
the lattice constant and $f=\Phi /\Phi_0$ is the magnetic frustration.
Henceforth, we rescale $H$ such that
$\omega_{H} \rightarrow H$. In the large $M$ (mean-field) treatment the bare
$T=0$ distance to the critical point $\delta$ is renormalized, and
the equation for the quasiparticle excitation gap $m$,
\beq\label{gap0}
m^2=\delta+\frac{H}{2}+\frac{UTH}{4\pi}\sum_{\omega_m, n}\frac{1}
{m^2+Hn+\omega_n^2},
\eeq
must be solved to obtain the effective Gaussian action.
Performing the summation over the Matsubara frequencies exactly
and employing the Poisson formula for the sum over `Landau' levels $n$,
we obtain (assuming $m \ll 1$)
\beq\label{gap}
\Delta &=& 2T\ln \left( 2\sinh\frac{m}{2T} \right)-
\frac{H}{4m}\coth\frac{m}{2T}-
\frac{H}{2} \int_0^\infty \frac{dx}{e^{2\pi x}-1}\cdot \nonumber\\
&&\frac{a_{-}\sinh(a_{+}/T)+a_{+}\sin(a_{-}/T)}{\sqrt{m^4 +H^2 x^2}
(\sinh^2(a_{+}/2T)+\sin^2(a_{-}/2T))},
\eeq
where
\beq
a_{\pm}=\frac{1}{\sqrt{2}} \left[ \sqrt{m^4 +H^2 x^2}\pm m^2
         \right] ^{1/2}\nonumber
\eeq
and $\Delta=(4\pi/ U)\delta+\Lambda$ ($\Lambda$ is the upper momentum cutoff)
is the renormalized distance to the zero-temperature critical point in the
absence of the field.
\begin{figure}
\begin{center}
\epsfig{file=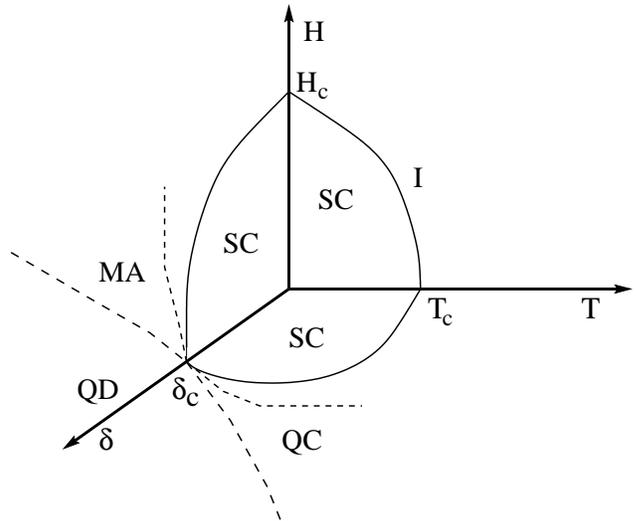, height=7cm}
\caption{Phase diagram for the field-tuned transition
in the presence of quantum fluctuations, $\delta$.
The zero-temperature quantum critical point studied
here corresponds to $\delta=\delta_c$.
Quantum disordered (QD), quantum critical (QC) and
magnetically activated (MA) regimes correspond to the limiting cases
$\Delta\gg\sqrt{H},T$, $T\gg\sqrt{H},\Delta$ and
$\sqrt{H}\gg T,\Delta$, respectively.}
\label{1}
\end{center}
\end{figure}

\begin{figure}
 \begin{center}\epsfig{file=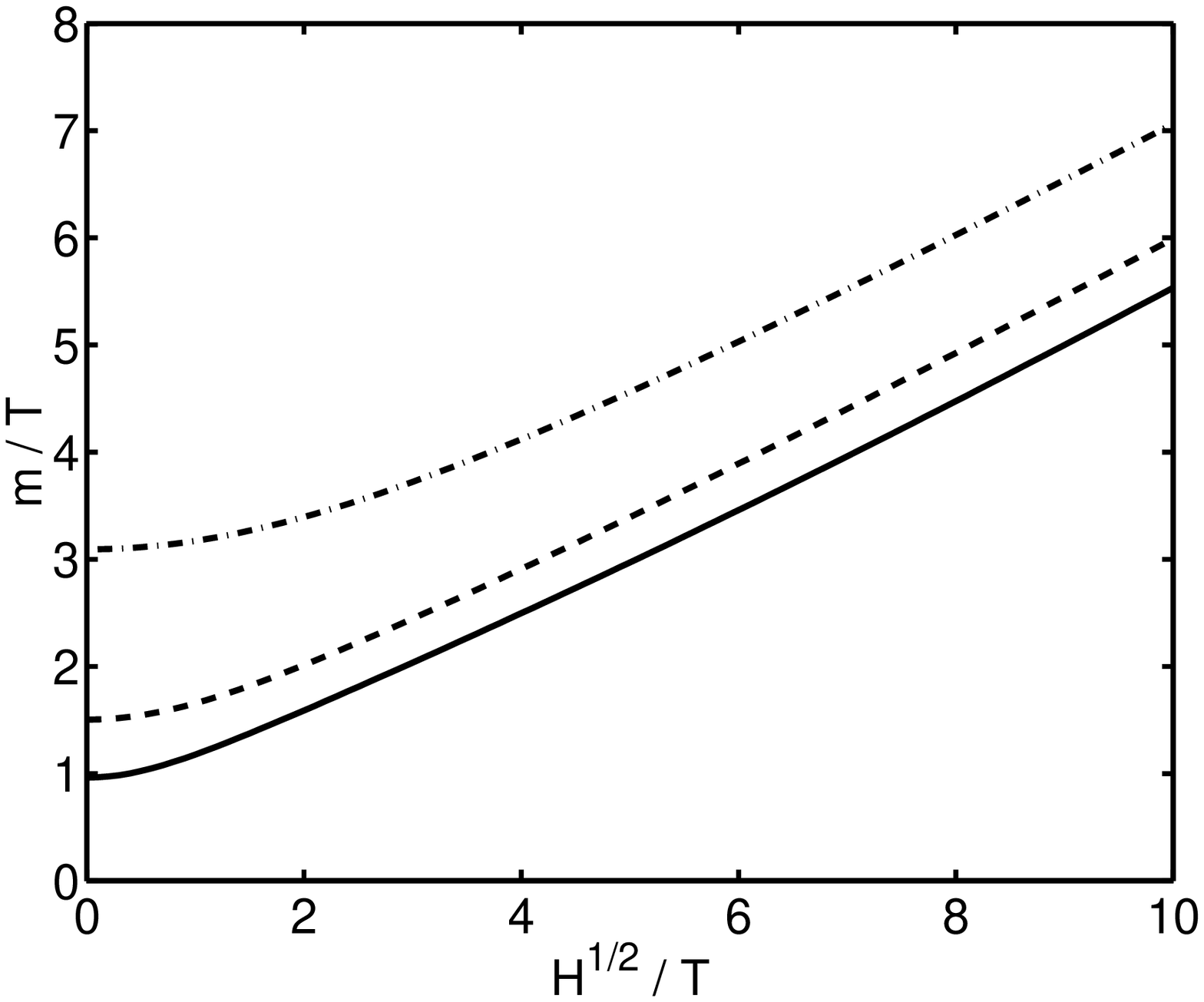,height=6.5cm}
 \begin{center}\epsfig{file=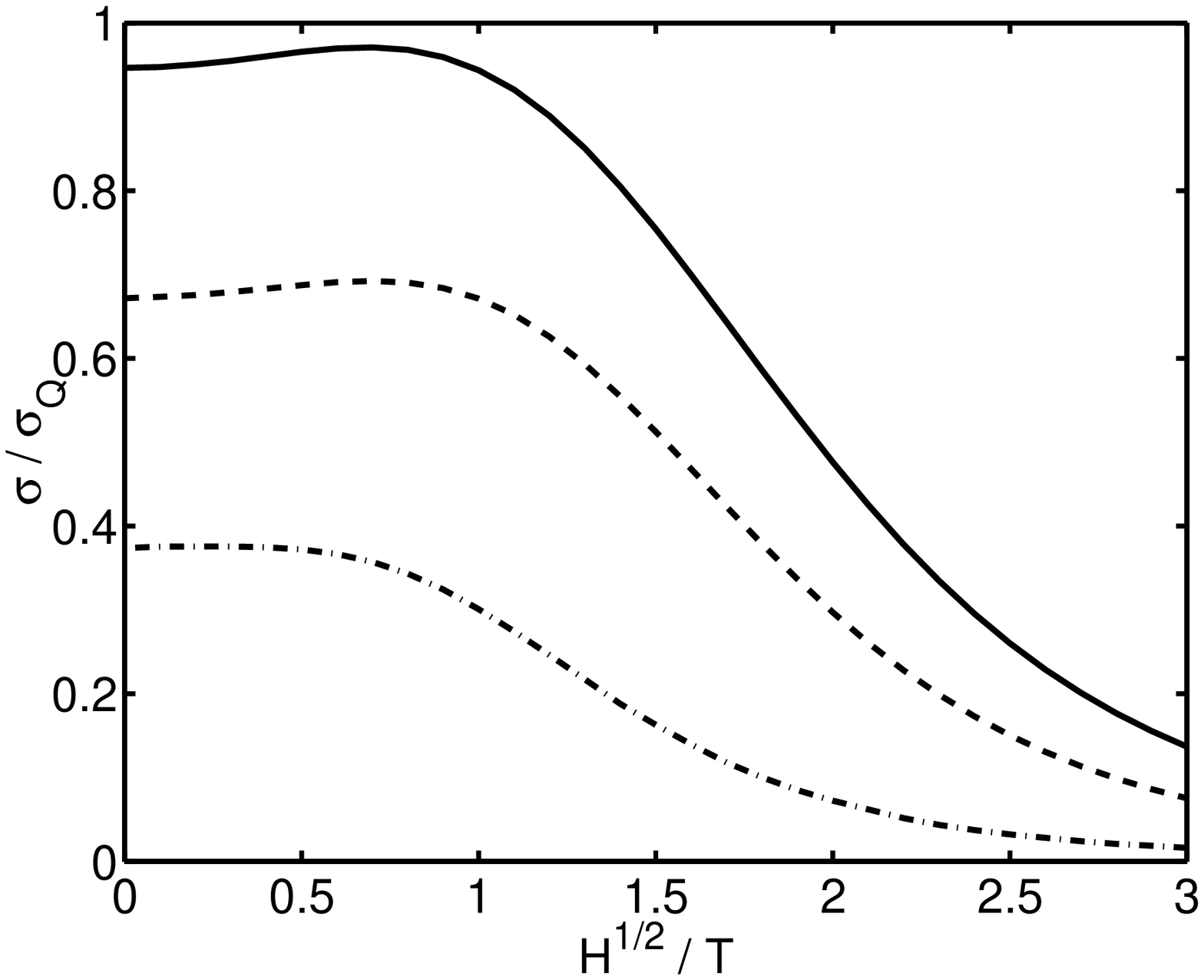,height=6.5cm}
 \caption{The dependence of $m/T$ and the corresponding conductivity
  $\sigma$ (in units of $\sigma_Q=4e^2/h$) on $\sqrt{H}/T$ for three values
  of the distance to the quantum critical point: $\Delta/T =0$
  (solid line), $\Delta/T =1.0$ (dashed line) and  $\Delta/T =3.0$
  (dashed-dotted line). Increasing $\Delta$ results in a crossover
from the quantum critical ($\Delta \ll T$) to the quantum disordered
($\Delta \gg T$) regime. }
\label{2}
\end{center}
\end{center}
\end{figure}

Three limiting regimes follow from Eq. (\ref{gap}).
1) If $T,\sqrt{H} \ll \Delta$,
then $m=\Delta$ and we are in the quantum disordered regime (QD).
This regime has been studied extensively recently, the key surprise
being the onset of a Bose metal phase\cite{denis2} once the Cooper pairs loose
their phase coherence.
2) The regime
$\sqrt{H},\Delta \ll T$ is quantum critical (QC) in which $m=\Theta T$
($ \Theta=2\ln((\sqrt{5}+1)/2)$).  Extensive
studies of this regime exist in the literature\cite{chubukov}.
In both of these regimes, the magnetic field effects are subdominant,
leaving the conclusions of the previous
 analyses\cite{denis2,chubukov} in tact. 3) Finally, when
$\sqrt{H} \gg T,\Delta$, the gap $m={\cal A} \sqrt{H}$ emerges,
where ${\cal A}=0.418945$ is the solution to the equation
\beq\label{A}
{\cal A}=\frac{1}{4{\cal A}}+\sqrt{\frac{1}{2}} \int_0^\infty
\frac{dx}{e^{2\pi x}-1} \left[ \frac{\sqrt{{\cal A}^4+x^2}-{\cal A}^2}
{{\cal A}^4+x^2} \right]^{1/2}.
\eeq
In this regime, which we refer to as magnetically activated (MA),
the correlation length, $m^{-1}$, exhibits universal behavior and
is determined (in the leading approximation) solely by the magnetic field $H$.
The energies of bosonic quasiparticles, which represent the low-energy
excitations close to the transition point, are quantized and occupy
the degenerate `Landau levels' with energies $\epsilon_n =\sqrt{Hn+m^2}$.
The $1/M$ correction is expected to modify the universal value of ${\cal A}$.
Its calculation is prohibitively difficult because the
coefficient in the quartic term of the GL action is
a complicated function of the energy levels $\epsilon_n$.
All of these regimes imply that the elementary excitations are
spin-wave like. This takes place only if $\Delta >0$. In the opposite
case, close enough to the finite $T$, finite $H$ transition,
transport is determined by the motion of vortices as illustrated
schematically in Fig. (\ref{1}).  The finite-size scaling considerations
suggest that the correlation length near the transition is a universal
function of two arguments: $\xi=T^{-1}f(\Delta /T ,\sqrt{H}/T)$ and
assumes the appropriate limiting forms if one of the parameters tends to
zero. The $M=\infty$ form of $m=\xi^{-1}$ as a function of
$\sqrt{H/T}$ obtained from the numerical solution of Eq. (\ref{gap})
for different values of $\Delta/T$ is presented on Fig. (\ref{2}).
As is evident, a crossover occurs to $\sqrt{H}$ dependence as the
field increases.

The linear conductivity (per flavor)
near the IST can be obtained from the GL action with the help
of the Kubo formula\cite{cha}
\beq\label{kubo}
\sigma_{\alpha\beta}(i\omega_n)=-\frac{\hbar}{\omega_n}\int d^2r\int d\tau
\frac{\delta^2\ln Z}{\delta \vec{A}_\alpha(\tau,\bf r)\delta \vec{A}_\beta(0)}
e^{i\omega_n\tau}.\nonumber
\eeq
While calculating the averages in the partition function
we expand $\psi$ using Eq. (\ref{psi}) which yields
for the longitudinal conductivity\cite{otterlo}:
\beq\label{sigma0}
\sigma (i\omega_{\nu})&=&\frac{(e^*)^2}{2h\omega_{\nu}} TH^2
\sum_{\omega_n} \sum_{n=0}^{\infty} (n+1) \left[ 2G(\omega_m ,n)\cdot
\right.\nonumber\\
&&\left.G(\omega_m, n+1)-G(\omega_m ,n)
G(\omega_m +\omega_{\nu}, n+1)\right.\nonumber\\
&&\left.- G(\omega_m ,n+1)G(\omega_m +\omega_{\nu}, n) \right]
\eeq
The experimentally observable finite temperature conductivity must
be calculated once dissipative mechanisms are taken into account
\cite{ds}.
Weak dissipation, ${\tilde \eta}=\eta +1/\tau$, is accounted for by 
introducing the term
${\tilde \eta} |\omega_m||\psi(\vec r,\omega_m)|^2$ into the free-energy
density\cite{cl}. Its inclusion changes the analytical properties of the Green
functions in Eq. (\ref{sigma0})\cite{denis}. Performing the
analytical continuation, one finds that the static conductivity is given by
\beq\label{sigma}
\sigma&=&\frac{(e^*)^2H^2}{4\pi h} \sum_{n=0}^{\infty} (n+1)\nonumber\\
&&\times\int_{-\infty}^{\infty} \frac{dx}{\sinh^2x}\frac{8{\tilde \eta}^2 T^2 x^2}
{(\epsilon_n -4T^2x^2)^2+4{\tilde \eta}^2 T^2 x^2} \nonumber\\
&&\times\cdot \frac{1}{(\epsilon_{n+1}-4T^2x^2)^2+4{\tilde \eta}^2 T^2 x^2}.
\eeq
The analysis of Eq. (\ref{sigma}) simplifies if the dissipation is weak
compared to the
quasiparticle energies, implying ${\tilde \eta} \ll m$. Asymptotic analytical
expressions are readily obtained for $m \gg T$, which define
QD and MA regimes. In these cases, there are two competitive
contributions to the integral over $x$.
One comes from the minima of the denominator at
$x_{n}^2 \approx (\epsilon_n^2-{\tilde \eta}^2 /2)/4T^2$ and
$x_{n+1}^2 \approx (\epsilon_{n+1}^2-{\tilde \eta}^2 /2)/4T^2$ and the second
from the region near $x=0$. Performing the approximate integration over $x$,
and analyzing subsequently the sums over $n$, we arrive at the following
limiting cases.

In the QD regime, $m=\Delta \gg \sqrt{H}$, and the conductivity is given by
$\sigma=\sigma^{(1)}+\sigma^{(2)}$, where
\beq\label{sigma1}
\sigma^{(1)}=\frac{4e^2}{\pi h} \left( \frac{\pi {\tilde \eta} T}{3 \Delta^2}
\right)^2
\eeq
and
\beq\label{sigma2}
\sigma^{(2)}=\frac{4e^2}{\pi h}
\left\{ \begin{array}{ll}
       \displaystyle\frac{2\pi T}{{\tilde \eta}} e^{-\Delta /T} & \quad
        H\ll \eta \Delta\\[4mm]
       \displaystyle\frac{8\eta T \Delta^2}{H^2} e^{-\Delta /T} & \quad
        {\tilde \eta} \Delta \ll H \ll \Delta T\\[4mm]
       \displaystyle\frac{\pi {\tilde \eta}}{T} e^{-\Delta /T} & \quad
        \Delta T \ll H \ll \Delta^2
       \end{array}\right..
\eeq
If we are in the MA regime with $\sqrt{H} \sim m$,
\beq\label{sigmama}
\sigma=\frac{4e^2}{h} \left[ \frac{{\tilde \eta}}{T}e^{-{\cal A}\sqrt{H}/T}
    +\frac{2\pi}{3}J \left( \frac{{\tilde \eta} T}{H} \right)^2 \right].
\eeq
In the formula above,
\beq
J=\sum_{n=0}^{\infty} \frac{n+1}{(n+1+{\cal A}^2)^2 (n+{\cal A}^2)^2}
 =23.913332\nonumber
\eeq
is a rather large numerical prefactor, arising from the
$M=\infty$ limit. These expressions describe qualitatively the crossover
from the QD to MA regime, provided the quasiparticles are well defined--
that is, dissipation is sufficiently weak. As
the field increases, the conductivity becomes directly
proportional to dissipation, as dictated by Eq. (\ref{sigma2}).
This can be attributed to the splitting of
the minima in the denominator of the general formula, Eq. (\ref{sigma}).

There are two sources of dissipation for which we must account.
Internal dissipation arises from
mutual scattering of quasiparticles,
leading to a finite quasiparticle scattering rate $1/\tau$.
External Ohmic dissipation $\eta$ may come from the resistive shunting of
superconducting grains in a system of fabricated
Josephson-junction arrays (JJA). The coefficient $\eta$ is then inversely
proportional to the resistivity of a shunting resistor, $R_g$\cite{wag},
which may be itself a function of temperature and magnetic field.
In addition to Ohmic dissipation, disorder can also serve as a dissipative
channel as has been shown for static disorder in the presence of d-wave pairing
\cite{herbut}.
In the QC regime, interactions alone lead to a large
inverse scattering time, $1/\tau \sim T \sim m$. Hence, quasiparticles are
poorly defined\cite{ds}, and small $\varepsilon=3-d$ or $1/M$
expansions are necessary to determine the conductivity in a controlled
way\cite{sachbook}. The implementation of this procedure leads to the universal
value for the conductivity close to $\sigma_Q=4e^2 /h$\cite{ds}.
The weak external dissipation $\eta \ll m \sim T$ is clearly
subdominant in this regime, except at low temperatures\cite{denis}.
On the contrary, in the QD regime, $1/\tau \sim Te^{-\Delta/T}$,
(up to logarithmic corrections),
as a result of the Boltzmann statistics of the quasiparticles.
In the absence of Ohmic dissipation, as follows from Eq. (\ref{sigma1})-
(\ref{sigma2}), metallic, rather than insulating behavior of the conductivity
obtains provided that $\sqrt{H}\ll \Delta$\cite{denis2}. The persistence of
the Bose metal phase for a finite range of magnetic field is possibly
relevant to the recent experiments of Mason and Kapitulnik\cite{mk}.

In the MA regime, quasiparticles
also obey Boltzmann statistics. Based on dimensionality
arguments similar to those valid for the QD regime, we suggest that in the
MA regime, $1/\tau \sim Te^{-{\cal A}\sqrt{H}/T}$, where the constant
of proportionality is determined by a numerical
coefficient that is of the first order in $1/M$.
The calculation of the corresponding lifetime is complicated because of the
discrete character of the quasiparticle spectrum.
However, as follows from Eq. (\ref{sigma2}), the conductivity becomes
directly, rather than inversely, proportional to dissipation for sufficiently
large $H$.  This means that inclusion of external dissipation $\eta$ and
determines only how fast
the conductivity decreases with increasing field. To illustrate this,
we resort to a semi-phenomenological calculation in which we set
$1/\tau =2.3300502 \cdot Te^{-m/T}$ and $\eta=0.1T$. The interpolation
formula for $1/\tau$ is chosen to satisfy the behavior in all regimes near
the critical point.
The numerical prefactor in $1/\tau$ is taken in such a way, that for
$\eta=0$, $\sigma \approx \sigma_Q$ in the QC regime.
The conductivity was then calculated numerically as a function
of $\sqrt{H}/T$ for different values of $\Delta /T$,
using Eqs. (\ref{gap}) and (\ref{sigma}).
The results that are depicted in Fig. (\ref{2}) support
rather rapid decrease of $\sigma$ with increasing $\sqrt{H}/T$
but their dependence on $\eta/T$ is relatively weak.
We conclude, then, based on the mean-field treatment, that the
conductivity decreases when the crossover to the MA regime occurs
even if $\eta=0$. This behavior is in agreement with experiments on the
magnetic field-tuned IST in disordered indium-oxide thin films\cite{hp}.
However, these results are in disagreement with the suggestion
of Kim and Weichman that the conductivity saturates with
increasing field\cite{km}.  This discrepancy can be explained
by the fact that for $\eta=0$ in the general universal scaling function,
\beq\label{scal}
\sigma=(4e^2/h)\Sigma(\omega /T, \Delta^{z} /T, \sqrt{H}/\Delta),
\eeq
applicable for the conductivity near the $d=2$ IST, the limits
$T=0$, $\omega \rightarrow 0$ and $\omega =0$, $T\rightarrow 0$
do not commute\cite{ds}. The former is the collisionless limit, apparently
considered in Ref. (14), in which the conductivity is likely to experience a
crossover between two universal values depending on the
ratio $\sqrt{H}/\Delta$\cite{otterlo}.  However, it
is the collision-dominated limit that
is experimentally relevant and hence the main subject of our treatment.
It is straightforward to verify that Eq.(\ref{sigma}) satisfies the general
scaling relation (\ref{scal}) with $\omega =0$ and $z=1$, 
provided that $\eta=0$ and 
the inverse scattering time $1/\tau$ obeys itself the scaling form 
 $1/\tau =T f(\Delta^{z} /T, \sqrt{H}/\Delta)$ in the critical region.  

The results above were obtained for low enough magnetic fields,
$\sqrt{H} \sim f \ll 1$ and hence, magnetic frustration
can be neglected. In translationally periodic systems such
as JJA, magnetic frustration
plays an important role giving rise to
peaks in the conductivity for integer values of frustration $f$\cite{webb}.
Moreover, the peaks corresponding to the rational values
$f=\frac{1}{2},\frac{1}{3}, \frac{3}{8}\cdots$
 may also be observed\cite{van der Zant}.
However, disorder disrupts the lattice periodicity
smearing the peaks in the conductivity
giving rise to a disappearance of frustration effects.
This situation takes place in the disorder-tuned IST in thin films.
Hence, we expect, that as long as Cooper pairs remain intact
and the transition is described by Eq.(\ref{action}), our
calculations provide a qualitative picture for the behavior of the
magneto-conductivity near the point of nominal the IST in thin
films including the observation of a Bose metal phase.

This work was funded by the DMR
of the NSF grant No. DMR98-96134.

\end{document}